\documentclass[a4paper,fleqn,usenatbib]{mnras}

\usepackage{newtxtext,newtxmath}

\usepackage[T1]{fontenc}
\usepackage{ae,aecompl}

\usepackage{graphicx,psfig}	
\usepackage{amsmath}	
\usepackage{amssymb}	
\usepackage{graphics}

\newcommand{\ltsima}{$\; \buildrel < \over \sim \;$}
\newcommand{\simlt}{\lower.5ex\hbox{\ltsima}} 
\newcommand{\gtsima}{$\; \buildrel > \over \sim \;$}
\newcommand{\simgt}{\lower.5ex\hbox{\gtsima}} 

\newcommand{\feka}{\mbox{Fe  K$\alpha$}}
\newcommand{\fekb}{\mbox{Fe  K$\beta$}}

\newcommand{\nika}{\mbox{Ni K$\alpha$}}
\newcommand{\xmm}{{\emph{XMM-Newton}}}

\newcommand{\lum}{erg~s$^{-1}$}
\newcommand{\flux}{{erg~cm$^{-2}$~s$^{-1}$}}
\newcommand{\nh}{cm$^{-2}$}
\newcommand{\nhsym}{N_{\mbox{\scriptsize H}}}

\newcommand{\chandra}{{\emph{Chandra}}}

\newcommand{\errUD}[2]{\ensuremath{^{+#1}_{-#2}}}

\newcommand{\fexxv}{Fe\,\textsc{xxv}}
\newcommand{\fexxvi}{Fe\,\textsc{xxvi}}
\newcommand{\suzaku}{{\emph{Suzaku}}}

\newcommand{\logxi}{erg cm s$^{-1}$}
\newcommand{\nustar}{{\emph{NuSTAR}}}

\newcommand{\sorg}{MCG-03-58-007}

\title[\sorg: a  new powerful   disk wind]{A new powerful and highly variable  disk  wind in an AGN-star forming galaxy, the case of \sorg}

\author[V. Braito et al. ]{V.  Braito$^{1,2}$\thanks{E-mail:valentina.braito@brera.inaf.it}, J . N. Reeves$^{2}$,  G.~A. Matzeu$^{1,3}$,  P. Severgnini$^{4}$,  L. Ballo$^{3}$, A. Caccianiga$^{4}$,
\newauthor
 S. Campana$^{1}$, C. Cicone$^{4}$, R. Della Ceca$^{4}$, T.~J. Turner $^{2,5}$ \\
$^{1}$INAF - Osservatorio Astronomico di Brera, Via Bianchi 46 I-23807 Merate (LC), Italy\\
$^2$Center for Space Science and Technology, University of Maryland Baltimore County, 1000 Hilltop Circle, Baltimore, MD 21250, USA\\
$^{3}$European Space Astronomy Centre (ESA/ESAC), E-28691 Villanueva de la Canada, Madrid, Spain\\
$^{4}$INAF - Osservatorio Astronomico di Brera, Via Brera 28, I-20121,  Milano, Italy\\
$^{5}$Department of Physics, University of Maryland Baltimore County, Baltimore, MD 21250, USA\\
 }

\date{Accepted XXX. Received YYY; in original form ZZZ}

\pubyear{2017}

\begin{document}
\label{firstpage}
\pagerange{\pageref{firstpage}--\pageref{lastpage}}
\maketitle

\begin{abstract}

We present the discovery of a new  candidate for a  fast disk wind, in the nearby Seyfert 2 galaxy \sorg. This wind is  discovered in a deep \suzaku\ observation  that was performed in 2010. Overall the X-ray  spectrum of \sorg\ is highly absorbed by a neutral column density of $\nhsym \sim 10^{23}$\,\nh,  in agreement with the optical classification as a type 2 AGN. In addition,   this observation unveiled the presence of  two deep absorption troughs   at $E=7.4\pm 0.1$\, keV and $E=8.5\pm 0.2$\,keV. If associated with blue-shifted  \fexxvi, these  features  can  be explained with  the presence of  two   highly ionised  (log $\xi/$\logxi$\sim 5.5$) and high column density  ($\nhsym \sim 5-8 \times 10^{23} $\,\nh) outflowing  absorbers with $v_{\rm out1}\sim -0.1c$ and $v_{\rm out2}\sim -0.2c$. The disk wind detected during this observation is most likely launched from within a few hundreds  gravitational radii  from the central black  and has a  kinetic output  that matches the prescription for significant feedback.  The presence of the lower velocity component of the disk wind is independently confirmed by the analysis of a follow-up     \xmm\ \& \nustar\   observation.  A  faster  ($v_{\rm out}\sim -0.35\,c$) component of the wind   is   also seen in this second observation.
 During this observation we also witnessed an occultation event lasting $\Delta t \sim 120 $ ksec, which we ascribe to an increase of the opacity of the disk wind ($\Delta \nhsym \sim 1.4 \times  10^{24}$\nh).  Our interpretation is that the slow zone  ($v_{\rm out}\sim -0.1\,c$) of the   wind is the most stable  but inhomogeneous  component, while the    faster zones could be associated with two  different inner streamlines of the wind.

\end{abstract}



\section{Introduction}
 Outflowing   ionised absorbers are a  common and  important occurrence of the central regions of   Active Galactic Nuclei (AGN).  These phenomena are not only important  in terms of the AGN  physics and the accretion process, but critically they  can play a key role in shaping  many galaxy properties, by providing the    feedback between the central super massive black hole (SMBH, $M_{\mathrm { BH}}=10^{6}-10^{9}$$M_\odot$) and  the host galaxy (\citealt{HopkinsElvis2010,DiMatteo2005}). X-ray observations of nearby bright   AGN unveiled the presence of  ionised outflowing absorbers   along the line of sight  in at least 50\% of radio quiet AGN (\citealt{Reynolds1997,Porquet2004,Blustin2005,Laha2014}). These ionised absorbers are observed over a wide range of ionisation ($\xi$\footnote{The ionisation parameter is defined as  $\xi=L_{\rm {ion}}/nR^2$,  where $L_{\rm ion}$ is the ionising luminosity in the 1-1000 Rydberg range, $R$ is the distance to the ionising source and $n$ is the electron density. The units of $\xi$ are \logxi.}), column density ($\nhsym$) and outflowing velocities.  Ionised absorbers that cover the lower end of the  ionisation and column density distributions (the so called ``warm absorbers") were revealed  in the late 1990's through the detection of absorption edges  in the soft X-ray spectra of nearby bright and unobscured  type 1 AGN (\citealt{Reynolds1997,George1998}).  
 With the advent of the modern high spectral resolution detectors on board \xmm\ and \chandra, a wealth of  blue-shifted absorption lines from different elements and ionisation levels  were detected in the soft X-ray band (\citealt{Crenshaw2003,Blustin2005,Kaastra2000,Kaspi2002,McKernan}). These soft X-ray absorption lines are   often blue-shifted with   velocities that range  from a few hundred km s$^{-1}$ to a few thousand km s$^{-1}$.

 The first evidence for the presence of more  massive and fast outflowing absorbers   came from the X-ray band too, through the detection of absorption features at rest-frame energies greater then 7 keV (e.g. \citealt{Chartas2002,Pounds2003,Reeves2003}). These absorption features are thought to be   blue-shifted K-shell absorption lines of highly ionised iron (\fexxv\   or \fexxvi).   Systematic investigations of the X-ray emission of nearby bright AGN showed that these highly-ionised  (log$\xi=3-6$), high column density    ($\nhsym \sim 10^{23}$\nh) and fast ($v>0.1 c$) winds    are present in  at least 40\% of the  X-ray bright, nearby  and radio-quiet AGN (\citealt{Tombesi2010,Tombesi2012,Gofford2013,Gofford2015}).  As the  velocities  can reach mildly relativistic values, their origin may be directly associated with the accretion process itself  (\citealt{King2003,King2010,King_Pounds2015}). The driving mechanism could be either radiation pressure  
 (\citealt{Proga2000,Proga2004,Sim2008,Sim2010}) or magneto-rotational forces (MHD models: \citealt{Kato2004,Kazanas2012,Fukumura2010,Fukumura2017}), or a combination of both.  Their outflow rates and   kinetic output can be huge and  match  or even exceed  the typical values necessary for significant feedback  on the host galaxy ($L/L_\mathrm{bol} \sim0.5-5$\%;  \citealt{HopkinsElvis2010,DiMatteo2005}). These disks winds can drive the massive molecular outflows seen on kpc-scales  thus simultaneously self-regulating the SMBH growth  and quenching the star formation, possibly leading to the observed  AGN-host galaxy relationships like the $M-\sigma$ relation (\citealt{Magorrian1998,Ferrarese2000,Gebhardt2000}).  Support for this scenario has recently come  from  the detection of   powerful X-ray  disk winds  in two Ultra Luminous Infrared Galaxies (ULIRGs),  where  massive large-scale molecular outflows are also present (Mrk 231; \citealt{Feruglio2015} and IRASF11119+3257; \citealt{Tombesi2015}). Furthermore, a large fast molecular outflow has been recently discovered in the lensed BAL QSO APM 08279+5255 (\citealt{Feruglio2017}), which has one of the fastest disk winds (\citealt{Saez2011}).
 
When multiple observations of ultra fast winds are available (e.g. PDS~456, {\citealt{Reeves2018a, Matzeu2017}; IRAS F11119+3257, \citealt{Tombesi2017}; PG1211+143, \citealt{pg1211} or APM 08279+5255, \citealt{Saez2011}), we often find that they vary either in column density, ionisation and/or  velocity.  The observed variations are  complex;  however,  such a complexity   is predicted  from   the disk wind models as the stream is not expected to be an homogeneous and constant flow  (\citealt{Proga2004,Giustini2012}). Thus,   depending on the observation, our line of sight may intercept different clumps or streamlines of the wind.   The best example  to date of a highly energetic wind is the one detected  in several X-ray observations of the  high luminosity QSO  PDS456 (\citealt{Reeves2003}), where the outflow velocity is as high as $v_{\rm{out}}\sim 0.3 c$ (\citealt{Reeves2009,Reeves2014,Gofford2014}).  The wind detected in PDS456 is characterised by one of the largest equivalent widths at  Fe-K known so far for these winds (from $\sim 100 $\,eV up to $\sim 500 $\,eV; \citealt{Reeves2014,Gofford2014,Matzeu2016,NardiniScience}, hereafter N15, and references therein). Furthermore, a  direct correlation between the outflow  velocity   and the intrinsic ionising  luminosity was recently reported for PDS456 (\citealt{Matzeu2017}) and IRAS 13224-3809 (\citealt{Pinto2018}). This suggests that  at least these  disk winds are predominantly radiatively driven: as the luminosity  and thus the radiation pressure increases, a faster wind is driven.  Examples of winds as powerful as the one detected in PDS 456 are still scarce; furthermore,   only in  a few cases we have multiple observations of the wind.    

\sorg\ is a  relatively bright and nearby  Seyfert 2 galaxy  ($z=0.031462$) and a Luminous Infrared Galaxy (LIRG, $L_{\rm IR}=1.7\times 10^{11}\,L_\odot$; \citealt{Rush1993}). Interestingly,  the profiles of the [O III]$\mathrm{\lambda 4959, \lambda5007}$~\AA{}  emission lines, in  the optical spectrum  from  the 6dF Galaxy Survey (6dFGS, Jones et al. 2009),  show   possible   broad blue wings ($FWHM\sim 900$\, km s$^{-1}$) blue-shifted by $\sim 500$\,km s$^{-1}$. Such   broad and blue-shifted lines could trace  the presence of a large scale galactic outflow. 
   \sorg\ was first selected  as a  candidate  to be a highly obscured AGN thanks to a short  \xmm\ ($<5$ ksec net exposure) observation (\citealt{Shu2008}). The  observed X-ray flux ($F_\mathrm{2-10\, keV}\sim  2 \times10^{-12}$ erg cm$^{-2}$ s$^{-1}$) was  indeed lower than the expectation from the bolometric    and infrared emission  and thus indicative of a highly obscured AGN  (\citealt{Severgnini2012}).
   
     Here we present the results of our follow-up observations performed with \suzaku\  (2010) and  \xmm\  \&  \nustar\ (2015), where we will show the detection of  a new  powerful  and highly variable disk wind, which strongly resembles the case of  PDS456.  As we will subsequently show, the wind was discovered in   \sorg\ thanks to a first  long  \suzaku\ observation performed in 2010. In contrast,  the  following simultaneous \xmm\  \&  \nustar\  observation caught an occultation event, which can be ascribed to a fast variation of the  disk wind. The paper is structured as follows: in \S 2  we describe the \suzaku\ and \xmm\  \&  \nustar\ observations and data reduction; in \S 3 we present  the spectral modelling focusing on the deep \suzaku\ observation and the discovery of  the occultation event in the 2015 observation. The discussion of the disk wind and  the physical interpretation of the  occultation event  are presented in \S 4.\ 
 Throughout the paper  we assume a concordance cosmology with  $H_0=70$  km s$^{-1}$ Mpc$^{-3}$, $\Omega_{\Lambda_{0}} =0.73$    and $\Omega_m$=0.27.

\begin{table*}
\caption{Summary of  the observations used: Observatory, Date, Instrument,  Elapsed  and Net exposure times.
 The  net exposure times    are after the screening of the cleaned event files. 
\label{table:log_observ}
}

\begin{tabular}{cclcc}
 \hline
 Mission &  Start-End Date (UT Time) & Instrument  & Elapsed Time (ks) &Exposure$_{\rm(net)}$ (ks)\\
 \hline

\suzaku\ & 	2010-06-03 16:50 $-$ 2010-06-05 21:27   &XIS  & 187.4  &  87.0\\

   \xmm\ &  2015-12-06 12:56 $-$ 2015-12-08 01:25 &   EPIC-pn&131.3 &  59.9\\
   \nustar\  &2015-12-06 10:36 $-$ 2015-12-09 17:21&FPMA &281.8 & 131.4\\
   
 \hline
\end{tabular}
\end{table*}

  \section{Observations and data reduction}

\subsection{\suzaku}
 \sorg\ was observed with \suzaku\  \citep{Mitsuda07} on 3$^{\rm rd}$  June 2010     for a
total  net exposure time of about 100 ksec (see Table~\ref{table:log_observ}).   We processed  the X-ray Imaging Spectrometer (XIS0, XIS3 and XIS1)  events using \textsc{HEASOFT} (version v6.21)  and 
the \suzaku\ reduction and analysis packages   applying the standard screening  criteria for the passage through
the South Atlantic Anomaly (SAA),  elevation angles and cut-off rigidity\footnote{The screening filters all  events  within
the   SAA  as well as  with an Earth elevation angle (ELV) $ < 5\ensuremath {{}^{\circ }}$ and  Earth
day-time elevation angles (DYE\_ELV) less than $ 20\ensuremath {{}^{\circ }}$. We also filtered the  data
within  256s of the SAA  from the XIS and within 500s of the SAA for the HXD. Cut-off
rigidity (COR) criteria of $ > 8 \,\mathrm{GV}$ for the HXD data and $ > 6 \,\mathrm{GV}$ for the XIS
were used.}. We extracted the XIS  source spectra  using a circular region of 1.9$'$ radius  centered on the source,  while the 
background spectra  were extracted from two circular regions of 1.5$'$ radius,   avoiding the calibration sources.  For each of the XIS, we generated the response and  ancillary response  files with the \textit{ftools} tasks \textit{xisrmfgen} and \textit{xissimarfgen}, respectively.  After checking that the XIS0 and XIS3 spectra were consistent, we combined them  in  a single  spectrum (hereafter XIS--FI). The combined FI (XIS1)  spectrum  was binned to 1024 channels and then grouped to a minimum of 50 (30) counts per bin.  For the spectral analysis we considered the XIS-FI spectrum over the 0.6--10  keV range while for the XIS1  we ignored the data above  8 keV,  since this CCD  is  optimised  for  the  soft  X-ray  band and has a higher background at higher energies.  For both  spectra we  also ignored    the 1.6--1.8 keV energy range because of calibration uncertainties.  We did not consider  in the analysis the  HXD-PIN data since \sorg\ is below the detection threshold.
 
 \subsection{\xmm}
 A simultaneous \xmm\  and \nustar\ observation of \sorg\ was performed in December 2015 (see Table 1). The \xmm-EPIC instruments operated in full  frame mode and with the thin filter applied. We processed and cleaned the \xmm\ data     using the Science Analysis Software  (SAS ver. 16.0.0)  and the resulting spectra were analysed  using the standard software packages  (FTOOLS ver. 6.22.1, XSPEC ver. 12.9; \citealt{xspecref}). 
 The EPIC data were first filtered for high background, which  affected this observation and reduced the net exposure time  to $\sim 60$ ksec,   
  $\sim 97$ ksec $\sim 98$ ksec for the pn, MOS1 and MOS2 respectively. 
  The EPIC-pn  source and background  spectra  were extracted  using a
circular region  with a   radius of $30''$  and    two circular regions  with a radius of $25''$, respectively. The EPIC-MOS1 and MOS2   source and background  spectra  were extracted  using a circular regions with a   radius of $25''$  and    two circular regions  with a radius of $30''$, respectively.   
We generated the response matrices and  the ancillary response files at the source position   using the SAS tasks \textit{arfgen} and \textit{rmfgen} and the latest calibration available.  For the scientific analysis    reported in this paper we  concentrated on the pn  data,  which have the highest signal to noise in the 2-10 keV band. The pn source spectrum was   binned to have at least 50 counts  in each energy bin.   The XMM-RGS and MOS spectra will   be presented in a companion paper that will discuss the soft X-ray emission and the broad band modelling (Matzeu et al.  in prep.), though the  MOS data were  checked for consistency with the pn.

 \subsection{\nustar}

\sorg\ was observed with the Nuclear Spectroscopic Telescope Array  (\nustar; \citealt{NUSTAR})    on  December 6$^\mathrm{th}$ 2015 for a total elapsed time of about $\sim280$ ksec, corresponding to a net exposure time of $\sim 130$ ksec. The observation was coordinated with \xmm\ and started $\sim 10$ ksec  before the \xmm\ observation, but extended nearly two days beyond the \xmm\ observation (see Table~\,\ref{table:log_observ}). We reduced the \nustar\ data following the standard procedure  using the \textsc{heasoft} task \textsc{nupipeline}   (version 0.4.5) of the \nustar\ Data Analysis Software  (\textsc{nustardas}, ver. 1.6.0). We  used the calibration files released with  the CALDB  version 20170222 and applied the standard screening criteria, where we filtered for the passages through the SAA  setting the mode    to  ``optimised"  in \textsc{nucalsaa}. For each of the Focal Plane Module (FPMA and FPMB)  the source spectra were extracted  from a circular region with a radius of $40''$,  while the background spectra were extracted  from two circular regions with a $45''$ radius located on the same detector. Light-curves in  different energy bands  were extracted from the same regions  using the \textsc{nuproducts} task. The FPMA and FPMB  background subtracted light curves  were then combined into a single one. From the inspection of  these light curves we defined three main  intervals (see \S 3.2.2),  characterised by a different   count rate, and created the corresponding Good Time Intervals  (GTI) files. These GTI files were then used to extract  source  and  background    spectra  and  the corresponding  response   files.   For the   time-resolved spectral analysis, the spectra were then binned to a minimum of 50 counts per bin and fitted over  the 3--65 keV  energy range.

\section{Spectral analysis}
All the spectral fits  were performed including  the Galactic absorption  in the direction of \sorg\ ($\nhsym=2.5\times 10^{20}$\,\nh, \citealt{Dickey}), which was modelled with the  Tuebingen - Boulder absorption model   (\textsc{tbabs} component in  \textit{XSPEC}, \citealt{Wilms2000}).  We   employed    $\chi^2$ statistics and   errors are quoted at the 90 per cent confidence level for one interesting parameter unless otherwise stated. All the parameters are quoted in the rest frame of \sorg\ ($z=0.0315$) and the velocities are all relativistically corrected\footnote{Both  velocities are    corrected with   the relativistic Doppler formula: $v/c=[(1+z_{\rm o})^2-1]/[(1+z_{\rm o})^2+1]$, where $z_{\rm o}$ is the measured rest-frame blueshift.}.

\subsection{The 2010 \suzaku\ observation}

 We first tested a   phenomenological model  typical of   a moderately obscured Seyfert 2  that is composed of an absorbed  primary power-law component and  a scattered component, with the same photon index ($\Gamma$).  This model provides a poor fit ($\chi^2=316.5/209$ d.o.f.), it requires a  rather steep photon index [$\Gamma> 2.9$, $\nhsym=(3.5\pm0.2)\times 10^{23}$\nh] and   fails to reproduce the overall spectral curvature.  
 
The inspection of the residuals  obtained by fixing the photon index to a  more typical value of $\Gamma=2$ (\citealt{ReevesTurner,Piconcelli2005,Caccianiga2004})  shows that the  steepness of the power-law component is not driven   by the soft X-ray emission as  is often seen in  other nearby bright AGN (\citealt{Turner1997,GuainazziBianchi2007}),  but mainly  by the presence  of two possible absorption features at $E>  7$ keV (see Fig.~\ref{fig:suzakuldra}).   In Fig.~\ref{fig:suzakuldra} we show the XIS spectra  against  this simple model (upper panel) and the relative residuals (lower panel). Since the   significance and strength of  absorption lines  strongly depend on   the intensity and slope of the primary emission, we first 
 proceeded to construct  a better description of the underlying continuum.  We   included in the model a  thermal emission component to account for the soft  residuals  (modelled with  \textsc{mekal}, \citealt{Mewe85}).  For completeness we also included  
 a narrow (consistent with the instrumental spectral resolution) Gaussian emission line component   to account for the  expected \feka\ emission line at $6.4$\,keV.
 \begin{figure} 
\begin{center}
\resizebox{0.49\textwidth}{!}{
\rotatebox{-90}{
\includegraphics{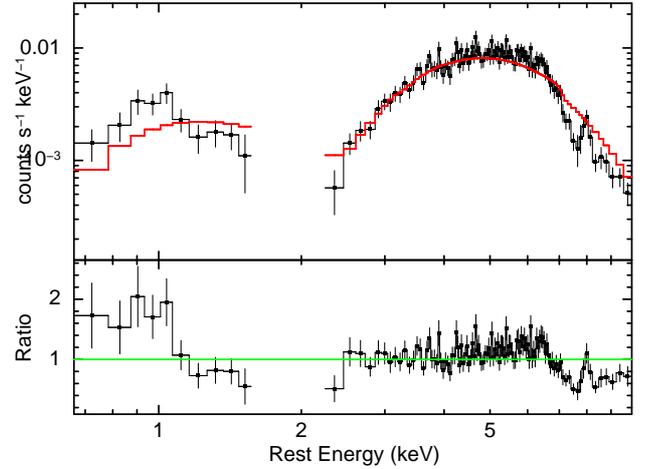}
}}
\caption{The upper panel shows the XIS-FI data (black data points) and     a simple  continuum    model  (shown in red), which is composed of an absorbed power-law component ($\Gamma=2 $,    $\nhsym \sim 2.5 \times 10^{23}$ \nh) and a scattered  soft  power-law  component.  The lower panel shows the data/model ratio  to this model.  Clear residuals are visible   in the soft X-ray band, due  to  possible thermal emission,  and at  the  iron K-shell energy band.   Besides a  weak   \feka\ emission line  at $\sim 6.4$ keV,   two strong absorption features are apparent at $\sim 7.4$ keV and at $\sim 8.5$ keV.    
 \label{fig:suzakuldra}
}
\end{center}
\end{figure}

 The addition of the \feka\ emission line improves the fit by $\Delta \chi^{2}/\rm{d.o.f.}=36.1/2$. The  energy  centroid is $E=6.40\pm0.05$\,keV and the  equivalent width is $EW=150\pm40$\,eV,   which is typical for mildly obscured Seyferts and  suggests the presence of  Compton-thick matter located out of the line of sight (\citealt{Murphy09}).  
 We therefore replaced the  Gaussian emission line with a cold  reflection component. For this component we adopted the    \textsc{pexmon} model (\citealt{pexmon}), which  includes  the  power law continuum reflected from neutral material as well as the \feka, \fekb, \nika\  emission lines and the \feka\ Compton shoulder.   For the reflection component we assumed    the same  $\Gamma$ of the primary  power-law component, an inclination angle of  $60$ degrees, a high energy cutoff of 100 keV. We fixed  the  amount  of reflection $R=\Omega/2\pi=1$ and allowed its normalisation to vary.  However, even with this more complex model the true shape of the primary emission remains highly  uncertain, because we  lack any information on the emission above 10 keV  and the    X-ray spectrum is highly curved and dominated by the  reprocessing  features (see Fig.~\ref{fig:suzakuldra}).

   We  thus decided   to  adopt the   best fit  photon index ($\Gamma=2.3$) that is  obtained later with the simultaneous \xmm\ \& \nustar\ observation, where \sorg\ was detected with high signal to noise (S/N) up  to 70 keV (see \S 3.2). The photon index was then  allowed to vary to within $\pm0.2$ of this value. From a statistical point of view this model    provides an  unacceptable fit ($\chi^2/\rm {d.o.f.}=329/206$) and  the residuals still show a broad absorption feature at $E\sim 7.4$ keV and a second feature at $E\sim 8.5$ keV  (see Fig.~\ref{fig:suzakura}, upper panel). As expected  $\Gamma$ hits    the upper limit of the range  ($\Gamma=2.5$),  even allowing the slope of the soft scattered component to be  independent from the shape of the hard component ($\chi^2=317.2/205$; $\Gamma_{\rm SOFT}> 4$).    
   
   We then included in the model two Gaussian absorption lines, assuming a common  line width. Since for  the width we   can   only place a lower limit of $0.25$ keV, we subsequently fixed it to this value. The addition of the  two Gaussian absorption lines   improves the fit for a $\Delta \chi^2=63.5$   and $\Delta \chi^2=24.4$ for the 7.4 keV and the  8.5\,keV  feature, respectively.      
Thus,  two absorption  lines are   both detected at a significance greater than 99.99\%. The lower energy line ($E=7.42\pm0.09$ keV) has an $EW=330\pm 70$\,eV,  while the higher energy one  ($E=8.51\pm 0.15$ keV) has an $EW=320\pm 110$\,eV.  
The most likely identifications  of  these  features are blue-shifted  $1s-2p$ transitions of \fexxv\ or \fexxvi\ (\citealt{Tombesi2012,Gofford2013}). If   associated with \fexxvi\ (which implies the smaller blue-shift) the  two features can  be explained with  the presence of  two   highly ionised and high column density absorbers outflowing with $v_1\sim -0.07c$ and $v_2\sim -0.2c$.  

 \begin{figure}
\begin{center}
\resizebox{0.48\textwidth}{!}{
\rotatebox{-90}{
\includegraphics{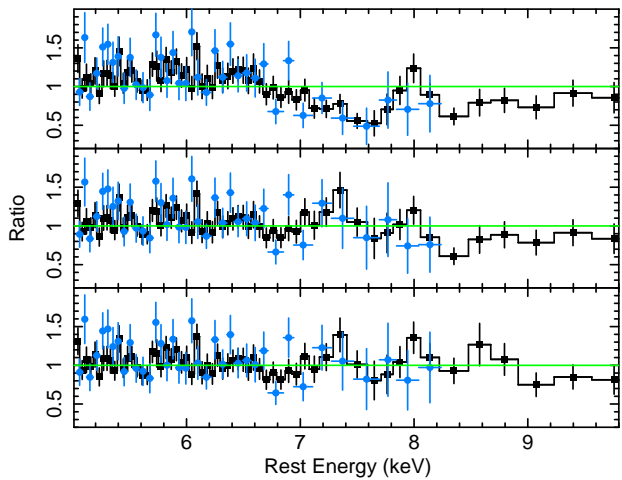}
}}
\caption{Zoom into the 5-10 keV residuals (XIS-FI data  black data points, XIS1  light-blue)  showing  the improvement of the  fit upon adding the ionised and outflowing absorbers. Upper panel: residuals to the   continuum model after the inclusion of  a cold reflection component ($\Gamma=2.5$).   A strong (EW $\sim 330$ eV) absorption line is present   at $E=7.4 \pm 0.1 $ keV, with a possible  additional absorption  feature at $E\sim 8.5\pm 0.1$ keV.  Middle panel: residuals to the model, where we included  one fast-outflowing ($v_{\rm out}/c=-0.076\pm 0.011$) ionised absorber; a  second absorption feature  is still present $E\sim 8.5$ keV. Lower panel: residuals  to the best fit model, with two outflowing zones,  where  the photon index is now similar to the  value obtained with the simultaneous \xmm\ \& \nustar\ observation  ($\Gamma=2.2\pm 0.2$).   \label{fig:suzakura}
}
\end{center}
\end{figure}
  
 To  self-consistently account for all    these features we replaced the two Gaussian absorption lines  with a  multiplicative grid of photoionsed absorbers generated with  the \textsc{xstar} photoionisation code (\citealt{xstar}). Since  the lines appear to be broad, we chose a grid that was generated   assuming a high turbulence velocity  ($v_{\rm turb}=10^{4}$\,km s$^{-1}$) and   optimised for high column density  ($\nhsym=5\times10^{22}$  - $2\times10^{24}$\nh) and high ionisation (log$\,\xi=3-8$) absorbers. The  grid was originally built  for PDS 456 (N15, {\citealt{Matzeu2016})  and assumes an  intrinsically steep ionising continuum, where the intrinsic photon index is $\Gamma=2.4$ (N15). This choice is  justified by    the   photon index   measured  for \sorg\ with \nustar, which     is as steep ($\Gamma\sim 2.3$; see \S 3.2) as in PDS456. 
 The inclusion of  one fast-outflowing ionised absorber  improves the fit by $\Delta \chi^2/\nu=68.9/3$.  As expected  from  the   parameters  of the Gaussian absorption line model, we found that the    absorber  is  outflowing with a velocity $v_{\rm out}/c=-0.075\pm 0.011$, is highly ionised log$\, \xi=5.5\errUD{0.5}{0.3}$ (e.g. consistent with the association  of the $\sim 7.4$\,keV feature with H-like iron) and  has a high column density $\nhsym=6\errUD{6}{2} \times10^{23}$ \nh.

  \begin{table} \caption{Summary of the   two phase disk wind model  for the \suzaku-2010 spectra. $^a$The ionisation of
the second zone was tied to the   ionisation of the lower velocity zone; if it is allowed to vary there is no improvement in the fit statistic. $^{b}$ The normalisation of the  scattered power-law component is highly degenerate with the  normalisation of the putative thermal emission component. The fluxes are corrected for the Galactic absorption, while the luminosities are intrinsic.  \label{tab:bestfit_neut}
}
 \begin{tabular}{llc}
\hline
 Model Component  &  Parameter  &  Value \\ 
 \hline
&&   \\
Primary Power-law &$\Gamma$&$2.2_{-0.2}^{+0.2}$ \\
& Norm. ($\times 10^{-3}$ ph keV$^{-1}$\,cm$^{-2}$)& $3.5_{-1.2}^{+1.7}$ \\

Scattered Component  &Norm. ($\times 10^{-6}$\,ph keV$^{-1}$\,cm$^{-2}$) $^{b}$&$8.9_{-5.8}^{+5.6}$\\
& & \\
Thermal emission   &kT (keV)&  $0.64\errUD{0.07}{0.07}$ \\
& Norm. ($\times 10^{-5}$ ph keV$^{-1}$\,cm$^{-2}$) & $2.2_{-0.5}^{+0.5}$ \\
&&\\
Neutral absorber&$N_\mathrm{H}$ ($ \times 10^{23}$ \nh)& $2.7\errUD{0.3}{0.2}$ \\
 &&\\
Zone 1   & $N_\mathrm{H1 }$\,($\times 10^{23}$ \nh )& $7.8\errUD{3.9}{2.5}$\\
              &log\,$ \xi_1$ &$5.54\errUD{0.29}{0.24}$ \\
               &$v_\mathrm{out1}/c$ &$-0.075\errUD{0.010}{0.011}$ \\
&&\\
Zone 2$^a$  & $N_\mathrm{H2}$\, ($\times 10^{23}$ \nh)& $5.4 \errUD{3.3}{3.0}$\\
              &log\,$ \xi_2$ &$5.54^t$ \\
&$v_\mathrm{out2}/c$ & $-0.20\errUD{0.02}{0.02}$ \\
&&\\
Reflection  & Norm. ($\times 10^{-4}$ ph keV$^{-1}$\,cm$^{-2}$) &$7.8\errUD{9.1}{5.4}$\\
     \\
    &$\chi^2/dof$&226.9/201\\
&$F_{(0.5-2)\mathrm {keV}}$(\flux) &$5.4\times 10^{-14}$\\
&$F_{(2-10)\mathrm {keV}}$ (\flux) &$2.0\times 10^{-12}$\\
&$L_{(0.5-2)\mathrm {keV}}$(\lum)  &$1.7\times 10^{43}$\\\
&$L_{(2-10)\mathrm {keV}}$ (\lum) &$1.4 \times 10^{43}$\\
 \hline
\end{tabular}
\end{table}

However, this ionised absorber is not able to account for the  second absorption feature detected at $E\sim 8.5$ keV (see Fig.~\ref{fig:suzakura}, middle panel), indeed  the  predicted    blue-shifted  \fexxvi\ Ly$\beta$  ($E_{\rm Lab}=8.25$ keV), which is included in the model,   would be too weak and    centered at $\sim 8.8$ keV.  We thus  included a second ionised absorber (modelled with the same grid) and assumed that the two zones share the same ionisation, but allowed the $\nhsym$ and velocity to be different. Here, two absorbers  are  required to produce the final best-fit   ($\chi^2=226.9/201$,  $\Delta \chi^2=21.4$ for 2 d.o.f.) where no strong residuals are left (see Fig.~\ref{fig:suzakura}, lower panel) and where  $\Gamma$ is now reasonably constrained ($\Gamma=2.2\pm0.2$).     In Fig.~\ref{fig:best_fit_xis.ps}  we show the resulting \suzaku\ spectrum with the best fit model overlaid, while the best fit parameters   are reported in Table 2.   Both  zones  of the ionised absorber are characterised by a high column density  ($N_\mathrm{H1 }=7.8\errUD{3.9}{2.5}\times 10^{23}$\,\nh\  and $N_\mathrm{H2 }=5.4 \errUD{3.3}{3.0}\times 10^{23}$\,\nh),   and are     outflowing with  $v_{\rm out1}=-0.075\pm0.01c$ and $v_{\rm out2}=-0.20\pm 0.02c$.    The high ionisation (log\,$\xi=5.5\pm0.3$),   outflow  velocity and  high velocity broadening are all indicative  of a disk wind launched from within few tens to hundreds of $R_\mathrm{s}$ from the central SMBH (see \S 4.1).

 \begin{figure}
\begin{center}
\resizebox{0.48\textwidth}{!}{
\rotatebox{-90}{
\includegraphics{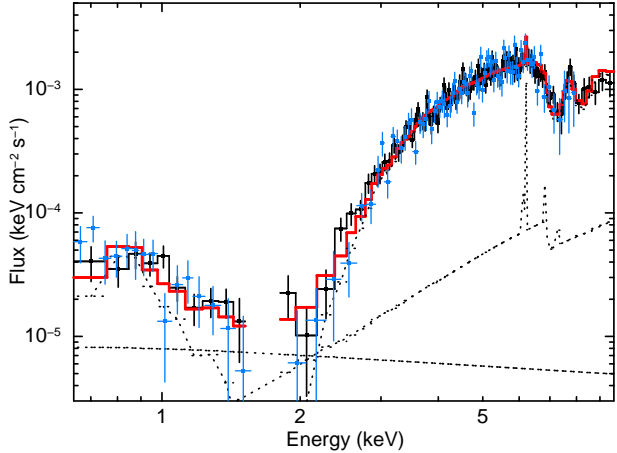}
}}
\caption{Suzaku data  (XIS-FI data  black data points, XIS1  light-blue) of \sorg\  when modelled with the two disk wind zones. The neutral absorber [$\nhsym = (2.7\pm 0.3) \times 10^{23}$\nh] is responsible for  most of the curvature that is present below 6 keV. The  imprints of the two  phase disk wind are the two strong  absorption features visible above  7 keV. A weak reflection component, which is most likely produced in the pc-scale putative torus, is present and  mainly account for the \feka\ emission line. To generate the plot, we created fluxed spectra   by unfolding the data  against a $\Gamma=2$ power law model and then we overlaid  the best-fit absorption model (red continuous line),   which is reported in Table~2; i.e. the data were not unfolded against the absorption model.
 \label{fig:best_fit_xis.ps}
}
\end{center}
\end{figure}

\subsection{Evidence for a variable absorber}
\subsubsection{The average  \xmm\ \& \nustar\ spectra}
  We then considered the   simultaneous  \xmm\ \& \nustar\ observation of \sorg\ performed five years later and  for simplicity we used only the EPIC-pn data for the \xmm\ observation. The observations started almost at the same time (see Table 1), but the duration of the \nustar\ one extended beyond  the \xmm\ orbit. We binned   both the \nustar\ FPMA/B spectra     to a minimum of 100  counts per bin and   the pn spectra    to a minimum of   50  counts per bin.   Although the \nustar\ observation  has a longer duration ($\sim300$ ksec) than the \xmm\ one, we first  jointly fitted the  pn  and  the \nustar\ spectra  extracted from  the whole  observation, allowing  for a cross normalisation  to account for variability and for instrumental calibration differences.  As per  the \suzaku\ spectra a  model   composed of an  absorbed power-law component and a scattered soft power-law component   (with the same $\Gamma$) provides a poor fit (with $\chi^2/{\rm d.o.f.}=1483.3/373$),  leaving  strong positive residuals above 10 keV and line-like residuals  both in the soft X-ray as well as at the  Fe K band.   In Fig.~\ref{compa_xmm_suzaku.ps} we show  a comparison   between  the \xmm\ \& \nustar\   and \suzaku\ data (the latter are included for comparison only).   The overall spectral shape and fluxes are similar, with a pronounced curvature between 2 and 6 keV, which is due to the fully covering and neutral absorber.      
  \begin{figure}
\begin{center}
\resizebox{0.48\textwidth}{!}{
\rotatebox{-90}{
\includegraphics{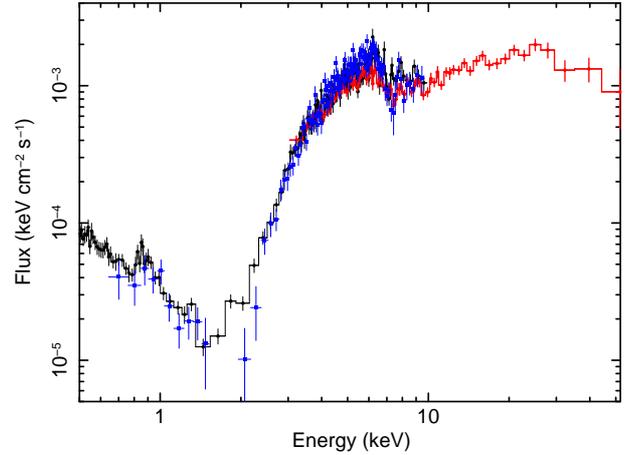}
}}
\caption{ Comparison between the spectra of \sorg\ collected during the \suzaku\ (blue data points) and  the simultaneous \xmm\ (black circles  points)  \& \nustar\ (red circles  points) observations. The fluxed spectra have been created  unfolding the data against a power law continuum   with  $\Gamma=2$. Note the similar level of the soft X-ray emission as well as the similar curvature in the 2-6 keV energy range, which is caused by the fully covering and neutral absorber ($\nhsym\sim 2\times 10^{23}$\,\nh).    
 \label{compa_xmm_suzaku.ps}
}
\end{center}
\end{figure}

  We  thus added a neutral  reflection component to the baseline continuum model. Since we   are not interested  here in the origin of the soft X-ray emission, which will be discussed  in a forthcoming paper,  we considered    the EPIC-pn spectrum   only above 2 keV and do not include in the model the scattered component.    We note that  the modelling of the soft X-ray emission does not impact the results obtained for the hard X-ray emission. Note that the soft X-ray spectrum is  constant below 2 keV and it is dominated by emission lines, which   likely originate  in a photoionised emitter (e.g. the Narrow Line Region gas)  with a small contribution from a  collisionally ionised gas (Mazteu et al. in prep). For the reflection component, we adopted the \textsc{pexmon} model and we allowed only its  normalisation  to  vary as we did for the \suzaku\ spectral analysis.
 Although the fit is still rather poor ($ \chi^2/{\rm d.o.f.}=414.1/278$), several things can be already noted. First of all the X-ray photon index is no longer  unusually steep  ($\Gamma\sim 2.23\pm 0.06$).  Secondly, the  cross-normalisations between  the \xmm\ and \nustar\ spectra are $C_{\rm FPMA}=0.83\pm0.03$  and  $C_{\rm FPMB}=0.86\pm0.03$,  which are  rather low for a simultaneous  observation. Furthermore the \nustar\ data do not perfectly overlap with the \xmm\ spectrum   (see Fig.~\ref{compa_xmm_suzaku.ps}).  Since the \nustar\ observation  has a longer duration (see Table~1),  we   investigated whether during this observation the X-ray emission of \sorg\ varied and  hence the averaged \nustar\ spectrum  is  fainter and harder  than the  pn spectrum.  \\

 \subsubsection{The \nustar\ lightcurves: evidence for an eclipsing event}
  In Fig.~\ref{fig:LC} we show the combined FPMA \& FPMB  light curves extracted  in the 3--6 keV (upper panel) and in  the 20--40 keV band (middle panel) together with the hardness ratios (defined as $HR= \rm CR_{20-40\,keV}/CR_{3-6\, keV}$; lower panel)  as a function of time. The light curves were extracted with a bin-size of 5814 sec corresponding to  the \nustar\ orbital period.  The two energy bands were chosen to track possible variations of the X-ray absorber  (3--6 keV) and/or the primary continuum (20--40 keV).  The inspection of    the   light curves    shows  that \sorg\   varied  during the \nustar\ observation   with a sharp drop in the 3--6 keV counts between 125-250 ksec, which is not accompanied by a similar drop in the hard X-ray counts.   This is suggestive of an eclipsing event that strongly resembles the occultations seen in  the light curves of NGC\,1365  (\citealt{Risaliti2009,Maiolino2010}),  which is often regarded as the prototype ``changing-look" AGN. Similarly to what we witnessed for NGC\,1365,   the  ingress of the absorber is rather sharp, while by  the end of the observation the source begins to  slowly uncover, as shown by the  recovery  in the 3--6 keV count rate after $\sim 250$\,ksec.   The hardness ratios  light curve shows that  \sorg\  becomes harder for almost half of the \nustar\ observation, with the HR increasing from $HR=CR_{\rm{20- 40\,keV}}/CR_{\rm{3-6\, keV}}=0.17\pm0.03$ to $HR=0.35\pm0.06$\footnote{The errors are  computed at the $3\sigma$ confidence level.}. 
  
  \begin{figure}
\begin{center}
\resizebox{0.45\textwidth}{!}{
\rotatebox{0}{
\includegraphics{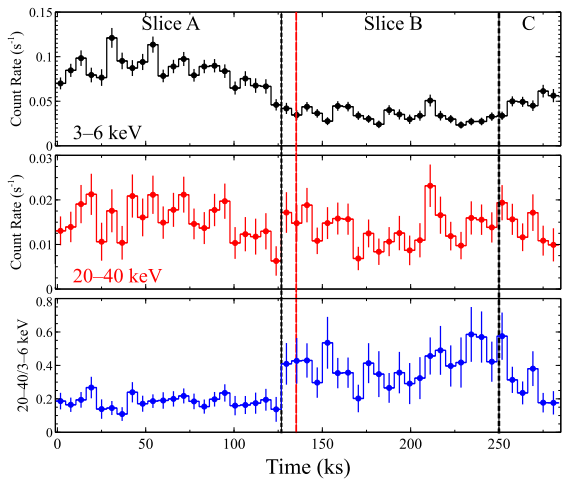}
}}
\caption{\nustar\   light curves extracted (with a bin-size  of 5814 sec) in   the 3--6 keV  (upper panel)  and the 20-40 keV band (middle panel). The lower panel shows the HR light curve defined as $CR_{\rm{20- 40\,keV}}/CR_{\rm{3-6\, keV}}$.    While there are only small fluctuations in the 20--40 keV light curve, the 3--6 keV light curve presents a sharp drop at around 125 ksec (marked with the black dashed line) into the observation and  a smooth increase at the end of the observation (at around 250 ksec, black dashed line). 
The HR light curve (lower panel) shows a sharp increase at 125 ksec, suggesting that we are witnessing an obscuration event. At 250 ksec the HR starts to decrease to almost the initial value.  The red dashed line marks the end of the \xmm\ observation at  around 140 ksec.
 \label{fig:LC}
}
\end{center}
\end{figure}

 \begin{figure}
\begin{center}
\resizebox{0.46\textwidth}{!}{
\rotatebox{-90}{
\includegraphics{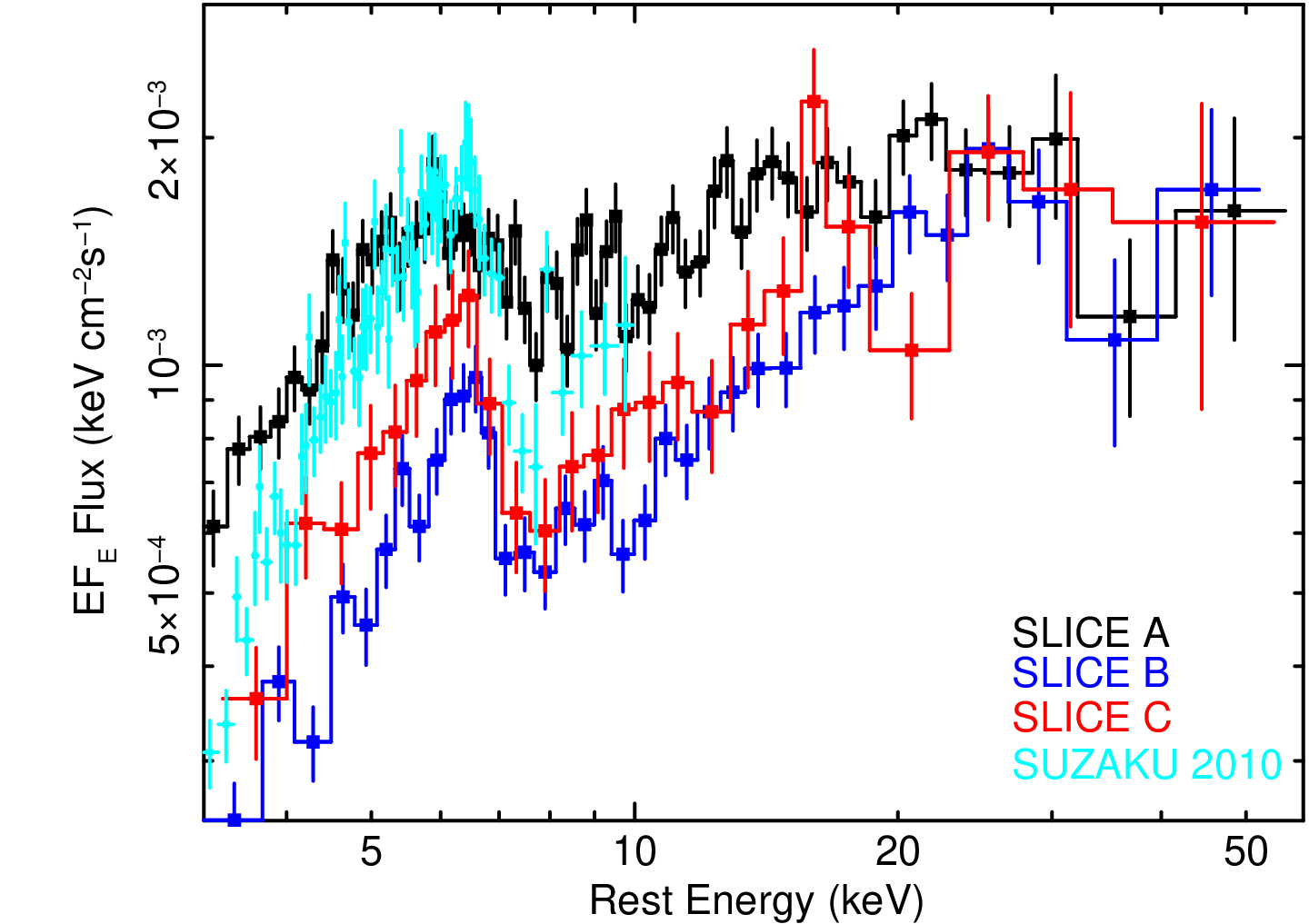}
}}
\caption{ Comparison between the \nustar\ spectra of \sorg, extracted following the HR light curve, and the \suzaku\ 2010 spectrum (shown in cyan).  The \nustar\ spectra  confirm that  during the \nustar\ pointing we   caught an occultation event. The first spectrum (sliceA, shown in black) is   typical  of a Compton thin state, while  the X-ray spectrum  of the second interval  (sliceB from 125-250 ksec, blue data points) is similar to a Compton thick AGN. Although limited by the short net  exposure time  of the third (after 250 ksec) spectrum  (shown in red), it is  apparent that at the end of the observation  \sorg\  begins  to uncover.   Note that the \suzaku\ spectrum is intermediate between the sliceA and sliceB spectra. To generate the  fluxed spectra,  we unfolded the data  against a power-law  continuum with  $\Gamma=2$. FPMA and FPMB data are  combined for  plotting purposes.
\label{fig:3slices}
}
\end{center}
\end{figure}

We therefore extracted  the \xmm\ and  \nustar\ spectra following the  hardness ratio light curve shown in Fig.~\ref{fig:LC}. Unfortunately when the variation occurred  we are left with less than 10 ksec of  the \xmm\ observation. Since  this latter EPIC-pn spectrum has a   low  S/N and since it covers only the beginning of the occultation,   it was not considered for the  time-resolved  spectral analysis.
 All the spectra were binned  to a minimum of 50  counts per bin.  The resulting \nustar\ spectra are shown in Fig.~\ref{fig:3slices}  and confirm  that during the observation we witnessed an occultation event.  At the beginning of the observation \sorg\ shows a    typical spectrum of a Compton thin AGN (black spectrum in Fig.~\ref{fig:3slices}). During the second half of the \nustar\ observation,  the X-ray spectrum  (blue spectrum in Fig.~\ref{fig:3slices}) is more curved  and  almost reminiscent of a  highly obscured  AGN.   The spectra also confirm that we caught both the ingress of the  obscuring cloud as well as the egress,  as    suggested by the hardness ratio light curve. Indeed, the third spectrum (shown in red in  Fig.~\ref{fig:3slices}) is less obscured  and \sorg\ starts  to uncover in the 3--6 keV energy range. However, to investigate in detail the origin of the spectral variability, we   considered   only  the first  (0--125 ksec) and the second  (125--250 ksec)  intervals  of the \nustar\ observation (hereafter  sliceA and  sliceB).  The quality of the spectra of the third interval is too low for a detailed spectral analysis, because we have less than 15 ksec of net exposure time  (per FPM  detector).   

\subsubsection{Time-sliced spectra}
  In order to derive a baseline continuum model we  first concentrated on sliceA. In Fig.~\ref{fig:Fe_ratio_nustar} (upper panel) we show the data/model residuals for the pn spectrum, which offers the best spectral resolution,  to a single  absorbed power-law ($\Gamma=2.19\pm0.06$) model over the 5--8 keV energy range ($\chi^2=400.8/276$).  Both the  \feka\ and \fekb\ emission lines  are clearly  visible as well as a weak absorption feature  at  the energy of  one of the absorption lines  detected in the \suzaku\ observation   ($E\sim 7.4$\,keV). Although the energy centroid of the absorption feature  ($E=7.4\pm0.1$\,keV) is consistent with the  lower energy absorption line detected in the \suzaku\ observation,   the feature  is  now significantly weaker  ($EW=64\pm 34$\,eV), suggesting some  variability of the disk wind.     We then included in the model a Gaussian absorption line and a  reflection component,  for which we adopted the \textsc{pexmon} model and allowed only its  normalisation  to  vary as we did for the \suzaku\ spectral analysis. 
  This model already provides a reasonable  description of the overall X-ray emission up to 65 keV ($ \chi^2/{\rm d.o.f.}=335.2/272$, $\Gamma=2.35\pm0.07$ and $\nhsym= (2.3\pm0.1)\times 10^{23}$\,\nh).

 The residuals of  sliceA and sliceB \nustar\ spectra against  this continuum model  (i.e. without any absorption feature) are shown in  Fig.~\ref{fig:slice1vsslice2} (upper panel),   where it emerges that   the \nustar\ spectrum collected during sliceB is more absorbed,  falling below the model up to 20 keV.  Conversely, above 20 keV, where the  spectra are dominated by the continuum emission,  the  sliceB spectrum is well predicted  by this model.  Thus,     if present,  variations of the primary emission  cannot account for the sharp drop in the 3--6 keV light curve. The  residuals of the more absorbed state are  also far from  featureless with two main absorbing structures between 7--12 keV. The lower energy  structure  is coincident with the  $E\sim 7.4 $ keV absorption feature detected in the \suzaku\ data, which appears to be much weaker in the sliceA  spectrum. 
To further investigate these absorbing  features, we then considered only   sliceB and fitted a single absorbed power-law component.  The ratio to this phenomenological model is shown in  the lower panel of Fig.~\ref{fig:slice1vsslice2}. A broad absorption trough  is present  blue-ward of the \feka\ emission line  at around $\sim 7.4$ keV as well as at    $\sim 10 $ keV and above.  We note that  the overall profile is highly reminiscent of  a P-Cygni profile from a wide angle outflow such as the one seen   in the X-ray spectra  of PDS456 (N15) and IRAS~11119+3257 (\citealt{Tombesi2017}), the prototype examples  of a wide-angle fast disk-wind.  

\begin{figure}
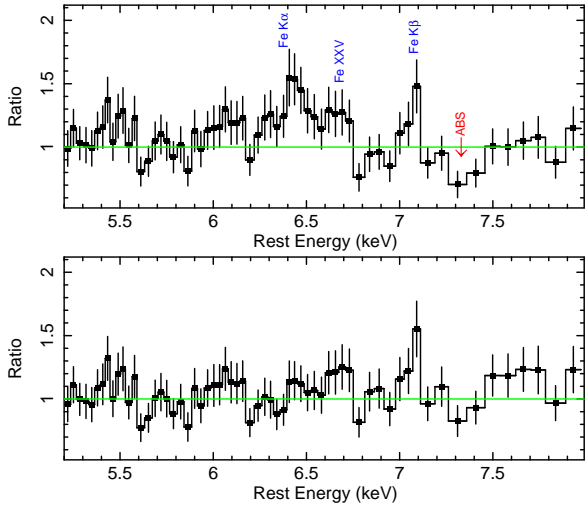

\begin{center}
\resizebox{0.45\textwidth}{!}{
\rotatebox{-90}{
\includegraphics{new_Fe_ratio.ps}
\includegraphics{final_fit_pn_ra.ps}
}}
\caption{Zoom into the 5-8 keV EPIC-pn residuals of sliceA. Upper panel:  residuals to the absorbed power-law model ($\Gamma=2.19\pm 0.06$).  The residuals show the presence of the \feka\ and \fekb\ emission lines and a possible excess at the expected energy of the \fexxv\ (labeled in blue). A weak absorption feature is present at  the energy of  one of the absorption lines  detected in the \suzaku\ observation ($\Delta \chi^2/\mathrm{d.o.f.}=11.5/3$; $EW=64\pm 34$\,eV; marked  in red).   Lower panel: EPIC-pn residuals  for the two-zone disk wind model (see \S 3.2.4). Only a weak residual left is at the energy of the Fe K$\beta$ emission line, which is underestimated with  the current pexmon model.
 \label{fig:Fe_ratio_nustar}
}
\end{center}
\end{figure}

Although the  high energies of these features   suggest that they   are most likely associated with blue-shifted \fexxvi\  resonance lines, and therefore with the presence of a highly ionised outflowing absorber, we  tested alternative scenarios.   We first investigated  if variations of the primary continuum could explain the spectral changes assuming either a constant reflection component or a constant reflection fraction. The former case represents a scenario where the reflection is produced in  distant material, such as the torus, which therefore does not respond to short time scale  variations of the primary continuum. The latter case mimics a scenario where the reflecting material is much closer in and immediately reverberates  the variations of the primary continuum.  Neither of the two scenarios provide  a satisfactory fit and both of them leaves strong residuals above 20 keV for sliceB. This is mainly caused by  a factor of 2-3 decrease of the normalisation  of the primary power-law component, which is  required to account for the lower level of 2-10 keV  emission seen during the second part of the observation. A strong decrease  of the intensity of the primary power-law component is found even upon allowing for  the neutral absorber to vary.  The residuals of the \nustar\ spectra  to the  model of the first scenario  including   a  variable neutral absorber ($\chi^2=470.1/345$\,d.o.f.) are shown  in Fig. \ref{fig:ra_evol2}  (upper panel). The model is unable to reproduce the  curvature of the sliceB spectrum leaving positive residuals above 15 keV.  \begin{figure}
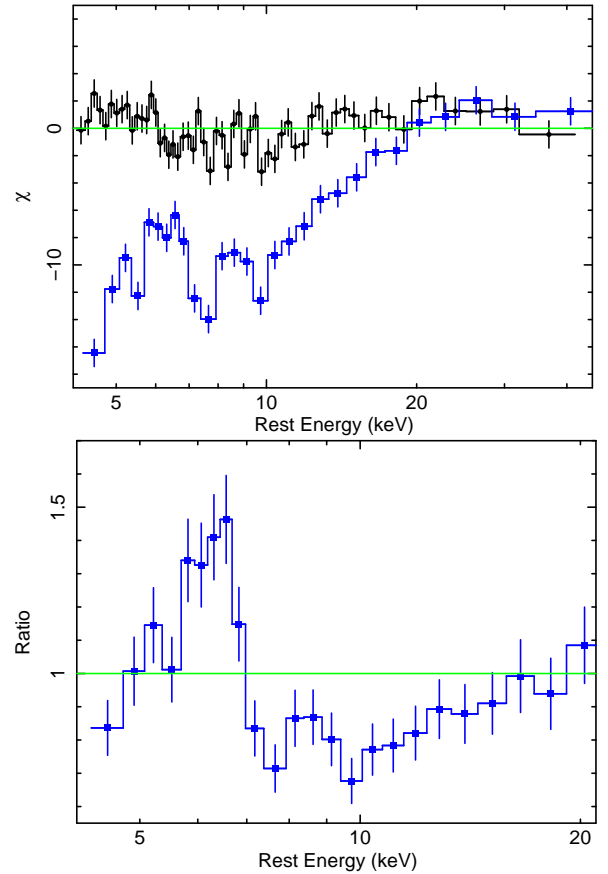

\begin{center}
\resizebox{0.46\textwidth}{!}{
\rotatebox{-90}{
\includegraphics{model1_over_slice2.ps}
\includegraphics{new_ra_slice2.ps}
}}
\caption{Upper panel: sliceA   (black data points) and sliceB (blue data points) residuals  to the best fit continuum model of sliceA. While only   weak absorption structures are  present in  the residuals of the first part of the observation, two deep absorption troughs  (at $E\sim 7.4$\, keV and $E\sim 10$\, keV) are evident in the  residuals of the more absorbed state (blue data points). 
 Lower panel: zoom into the 4--20 keV    data/model ratio between  the sliceB spectrum and an absorbed power-law continuum model. Two deep absorption features  are  evident in the data. The  profile of the lower energy structure is highly reminiscent of a  broad P-Cygni  profile. FPMA and FPMB data are  combined and rebinned  to 120 counts per bin for plotting purposes.
\label{fig:slice1vsslice2}
}
\end{center}
\end{figure}

We note  that, since all these variable absorption models  are neutral,   none of them is able to account for the absorption features that are evident  in the   7--12 keV  residuals of sliceB. We  then added to the baseline continuum model  two  Gaussian absorption lines, with variable depth between the two slices. As the width of both the  Gaussian absorption lines   is poorly constrained, they were fixed it to $0.1$ keV. The  fit improves  by $\Delta \chi^2=14.8$ and $\Delta \chi^2=27.1$ for the lower and higher energy features. The  absorption lines are  detected at $E=7.4\pm 0.1$\,keV and $E=10.2 \pm 0.1$\,keV, respectively.   Their  equivalent widths,   measured against the primary power-law component are $EW_{\rm 7.4\,keV}=70 \pm 36$ eV  and  $EW_{\rm10\, keV}=160\pm 100$\, eV for slice A and moderately stronger in sliceB with   $EW_{\rm 7.4\, keV}=170\pm 70$ eV and $EW_{\rm10\, keV}=240\pm 130$ eV.

\subsubsection{Modelling with a variable disk wind}
 A plausible  scenario is that  the variable absorber is   the disk wind that was detected in the \suzaku\ observation. We thus    included in the model  the same  multiplicative grid of photoionised absorbers   adopted for the \suzaku\ spectra.    We constrained the    ionisation and outflow velocity of this ionised absorber to be the same  between the slices and allowed only the column density  to vary between the two  slices. We did not allow the velocity to vary as the absorption features are seen at the same energy.    We also allowed the neutral absorber and the  normalisation of the primary power-law component to vary, while we assumed a constant reflection component.  Upon adding this ionised absorber the fit improves  by $\Delta\chi^2=60.9$ (for 4 d.o.f.). As expected the main driver of the spectral variability  is an increase of the column density of this  highly ionised (log\,$\xi=4.1\pm0.1$) and  outflowing ($v_{\rm out}=-0.12\pm0.01c$) absorber from $N_{\rm HA}=1.5\errUD{0.8}{0.6}\times 10^{23}$\nh\ to $N_{\rm HB}=(1.5\pm 0.2)\times 10^{24}$\nh. We note that a worse fit is obtained if we allow only the ionisation to vary instead of the $\nhsym$ ($\Delta\chi^2=18.4$). The neutral absorber  is  poorly constrained in sliceB, because  with   the \nustar\ data    alone we can place only an upper limit of $N_{\rm HB}<1.1\times 10^{23}$\nh,  while in sliceA we derive $N_{\rm HA}= (2.1\pm 0.2)\times 10^{23}$ \nh.  However, if we assume a constant neutral absorber,      a similar increase in $\nhsym$ is observed between sliceA and slice B   ($\Delta \nhsym\sim 8 \times 10^{23}$\,\nh).This model can now   better   reproduce the different spectral curvature of sliceA  and sliceB as well as the lower energy absorption structure at the Fe-K energy band. However, as expected from the energies of the two absorption features,  this absorber cannot  account  for the  structure  at $\sim 10 $ keV, because  this feature is too deep to be explained with  the corresponding  higher order \fexxv\ or \fexxvi\ absorption lines.       
  We thus included in the model a second ionised and outflowing absorber,  allowing only  its column density  to vary between the slices. We also allowed this absorber to have a different ionisation of the first absorbing zone.  The fit improves   by   $\Delta \chi^2=19.3$ for 4 d.o.f.,  which indicates that this additional zone is required at  confidence level  $>99.9$\% according to the  F-test.    Since there is no evidence for variability of    the   column density  of the faster zone ($N_{\rm H2A}=8.2\errUD{5.2}{3.2}\times 10^{23}$\nh\,$N_{\rm H2B}>4\times 10^{23}$\,\nh), we  then tied its $\nhsym$  between the two slices ($N_{\rm H2}=8.1\errUD{4.8}{3.1}\times 10^{23}$\nh; $\chi^2/ \nu=389.8/338$).     This second and faster  [$v_{\rm out2}=(-0.35\pm 0.02)\,c$] zone is  characterised by a higher ionisation   than the first zone log\,$\xi_2=6.2^{+0.2}_{-1.4}$.    
 
    We then tested if the observed variability could be instead explained   with the presence of a variable and neutral partial covering absorber. We thus tied the column densities of the outflowing ionised absorbers and added a partial covering absorber, allowing the covering fraction to vary. From a statistical point of view  this model returned an acceptable  fit $\chi^2/\nu=379.4/337$. However,  the $\nhsym$ of this absorber is found to be $>3.5\times10^{24}$\,\nh\  and almost fully covering ($\sim 99$\%)  during sliceB. This implies a rather extreme  and most likely  unphysical scenario. Indeed, once  the  Compton scattering is taken into account   for such a high $\nhsym$,  the  intrinsic 2--10 keV luminosity of \sorg\  would be of the order of $\sim 4\times 10^{44}$\,\lum\  and  thus  too  high  when compared to our estimates of the bolometric luminosity ($L_{\rm{bol}}=1.6-3.4\times 10^{45}$\,\lum, see \S 4).
   
    The relatively low  ionisation  of zone 1  (log$\xi=4.1\pm0.1$) implies that the $\sim 7.4$\, keV absorption line is  associated with  \fexxv\ instead of \fexxvi. 
  This explains why we now derive a higher outflowing velocity  ($v_{\rm out1}\sim-0.12\,c$) than  with the \suzaku\ spectra ($v_{\rm out}\sim-0.075\,c$), which also displayed an absorption feature at $\sim7.4$\,keV.  Note that  if we assume for this zone the same ionisation derived with the \suzaku\ spectra the fit is worse by $\Delta \chi^2=20.2$.  A possibility is that the cloud responsible for the occultation event is a denser clump with a lower ionisation.  To test this  scenario, we allowed also the ionisation of zone1 to vary between the slices. We also allowed the outflowing velocities to  be different in order to  adjust for the different ionisation. We  found that, while in sliceA the ionisation could be similar to the \suzaku\ observation (log\,$\xi_{\rm A}=5.5_{-1.1}^{+1.2}$; $v_\mathrm{out1}=-0.07\errUD{0.03}{0.02}$),  in the second part of the observation this zone  has an ionisation of log\,$\xi=4.1_{-0.09}^{+0.09}$  and  it is outflowing with $v_{\rm out}=-0.13\pm0.02\,c$. 
 This fit also confirms that the obscuring cloud has a column density of $N_{\rm H1B}=1.4_{-0.4}^{+0.2}\times 10^{24}$\,\nh.  The parameters for zone2 are almost identical  to the previous model, because  they do not  depend on the lower velocity zone ($N_{\rm H2}=9.4\errUD{5.5}{3.4}\times 10^{23}$\nh, log\,$\xi=6.2_{-1.5}^{+0.2}$,  $v_{\rm out2}=-0.35\pm0.02c$). 
 
 Here, two  highly ionised  and variable absorbers  are required to  have an acceptable fit for  the time-sliced spectra.   The main contribution to   zone 1 is the absorption structure at $\sim 7.4$\,keV,  as well as the  spectral curvature below 10 keV, which  becomes more  pronounced  during the second part of the observation and drives  the increase in  the column density. Conversely,   zone 2, given its higher ionisation, does not imprint  strong curvature to the  observed spectra  and mainly accounts for the higher energy absorption structure seen at $\sim 10$ keV (see Fig.~ \ref{fig:ra_evol2}, upper panel). The  best fit parameters of this model are listed in Table 3, while the resulting   residuals and spectra are  shown in Fig.~ \ref{fig:ra_evol2} (lower panel) and in  Fig.~\ref{fig:eeuf_final}, respectively.    The final fit statistic   ($\chi^2/\nu=382.7/336=1.14$)  is now good and no strong residuals are present,  with the exception of the weak absorption feature at $\sim 6.7-6.9$ keV, which is visible  only in the  pn data of sliceA. This feature could be a signature for   a lower velocity and lower ionisation phase of the wind. We thus included in the model a third ionised absorber, which was  also modeled with a  multiplicative grid of photoionised absorbers generated with  the \textsc{xstar} photoionisation code. We adopted a grid  that has   a lower turbulence velocity ($v_{\rm turb}=300$\,km s$^{-1}$), because the absorption line seen in the pn appears to be narrow. The fit only marginally improves $\Delta\chi^2/\nu=10.7/3$, but  confirms that an additional ionised absorber could be present; however,  its parameters are poorly constrained ($\nhsym \sim 1\times10^{23}$\,\nh, log\,$\xi\sim 2.5$  and  $v_{\rm out}=7000\pm 3500$ \,km s$^{-1}$).  Note that the inclusion of this latter absorber does not affect the main parameters of the two  fast zones of the disk wind.  The Fe-K band residuals to this model are shown in the   lower panel of Fig.~\ref{fig:Fe_ratio_nustar}.

  \begin{table} 
  \caption{Summary of the   two phase disk wind model applied to sliceA and sliceB  (\xmm\ \& \nustar\ 2015 observation). 
$^a$: The outflow velocity of this zone was allowed to vary to adjust for the different ionisation.
$^b$: the uncertainties on the column densities are determined for  the best fit ionisation as they are highly degenerate with log$\xi$. 
$^c$: the normalisation units are   $10^{-3}$ ph keV$^{-1}$\,cm$^{-2}$.
$^t$: denotes parameter was tied. \label{tab:2zones_nustar}
 }
   \begin{tabular}{llcc}
\hline
 Model Component  &  Parameter  &  SliceA & SliceB \\ 
 \hline
&&   \\
Primary Power-law &$\Gamma$ & $2.28_{-0.07}^{+0.07}$  & $2.28{^t}$\\
& Norm.$^c$ & $3.0_{-0.4}^{+0.4}$ & $2.3_{-0.4}^{+0.5}$ \\

&&\\
Neutral absorber &$N_\mathrm{H}(\times 10^{23}$ \nh)& $2.3\errUD{0.1}{0.1}$  &$<1.1$ \\
 &&\\
Zone 1$^a$     & $N_\mathrm{H1 } $($\times 10^{23}$ \nh)& $1.9\errUD{1.1}{0.7}$& $13.8\errUD{2.3}{3.5}$ \\
              &$log \xi_1$&$5.5\errUD{1.2}{1.1}$ & $4.09\errUD{0.09}{0.09}$  \\
               &$v_\mathrm{out1}/c$&$-0.07\errUD{0.03}{0.02}$  &$-0.13\errUD{0.02}{0.02}$\\
&&&\\
Zone 2$^b$    & $N_\mathrm{H2 }$ ($\times 10^{23}$ \nh)& $9.4\errUD{5.5}{3.4}$ & $9.4^t$\\
              &$log \xi_2$&$6.2\errUD{0.2}{1.5}$ & $6.2^t$  \\
&$v_\mathrm{out2}/c$ & $-0.35\errUD{0.02}{0.02}$ &$-0.35^t$ \\
&&\\
Reflection  &  Norm.$^c$&$1.8\errUD{0.7}{0.6}$&$1.8^t$\\
     \\
   \hline
\end{tabular}
\end{table}

\begin{figure}
\begin{center}
\resizebox{0.46\textwidth}{!}{
\rotatebox{-90}{
\includegraphics{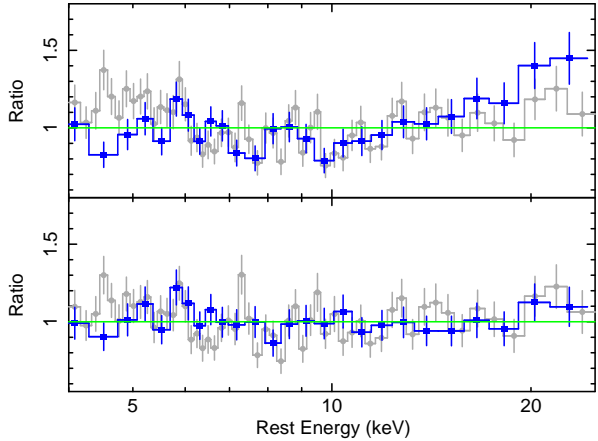}
}}
\caption{Zoom into  the 4-25 keV range of the data/model residuals to the different variable absorber models    fitted to  the sliceA (grey data points) and sliceB (blue data points) spectra. The FPMA and FPMB spectra  were co-added for plotting purposes. Upper panel: residuals to the model, where  only   the intensity of the primary power-law component and the amount of   neutral  absorption were allowed to vary. The model is unable to reproduce the  curvature of the sliceB spectrum. Two main absorption structures are also visible at around $\sim 7.4 $ keV (mainly in sliceB) and 10 keV. Lower panel:  residuals to the two-zone disk wind model.  After the inclusion of this ionised absorber, we can also reproduce   the overall spectral curvature. \label{fig:ra_evol2}
}
\end{center}
\end{figure}

\begin{figure}
\begin{center}
\resizebox{0.46\textwidth}{!}{
\rotatebox{-90}{
\includegraphics{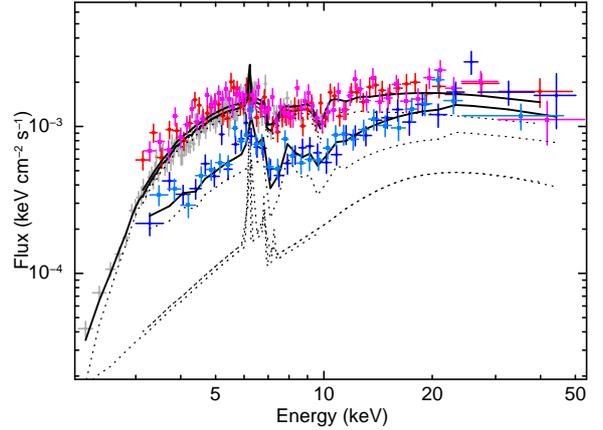}
}}
\caption{XMM \& \nustar\  sliceA and sliceB spectra. The  spectra of the first part of the observations  (pn grey,     \nustar\  FPMA  red  and  FPMB magenta data points)  and the second part of the observation (\nustar\  FPMA blue and  FPMB  light blue data points) were fitted with a   variable  disk-wind model. The  baseline  continuum model is   composed of an absorbed power-law component and reflected component.  The $\nhsym$ of the  neutral absorber and  of the low-velocity    ionised  absorber (zone1) are  allowed to vary. Zone1 is  the main driver for the higher  curvature  observed below 20 keV   during  the second half of the observation, with a variation in the amount of absorption  of $\Delta \nhsym \sim 1.4\times 10^{24}$\,\nh. The fluxed spectra were generated as in Fig.~3.
\label{fig:eeuf_final}
}
\end{center}
\end{figure}

\section{Discussion and conclusion}
  We have presented  the discovery of a new and highly variable disk wind. The disk wind has been discovered thanks to our long \suzaku\ observation, where  two zones of a fast ($v_{\rm out1}=-0.075\pm 0.01\,c $ and $v_{\rm out2}= -0.2\pm 0.02\,c$) highly ionised absorber were revealed through the presence of two deep absorption features. 
  This    places \sorg\ among the fastest  and potentially most powerful winds detected.   Indeed, in only a handful of the disk winds discovered so far  the velocity exceeds $0.2\,c$ (\citealt{Gofford2015,Tombesi2012}). In the follow up observation, performed simultaneously by  \xmm\  \& \nustar, we discovered a fast occultation event, which we ascribe to an increase of the $\nhsym$ of the slower  zone of the disk wind from $\nhsym=1.9\errUD{1.1}{0.7}\times 10^{23}$\,\nh\ to $\nhsym = 1.4\errUD{0.2}{0.4}\times 10^{24}$\,\nh.      The \suzaku\ observation has a duration of almost 200 ksec  and  shows strong variations, up to a factor of $\sim 2$,  in the 
   2--10 keV X-ray band. However, without the bandpass above 10 keV, we cannot decouple whether      the 
  variations are due to    continuum  or column density changes. However, it is possible that during 
the  \suzaku\  observation our line of sight  intercepted another clump  of the wind with a $\nhsym$ that is intermediate between the $\nhsym$ measured in sliceA and sliceB. Indeed, as shown in Fig.~\ref{fig:3slices}, the \suzaku-spectrum falls  between the \nustar\ spectra.  
  To  explore the properties of the  disk wind detected in \sorg, within the context of  its possible  impact on the  host galaxy, we first need to derive a first order estimate of its mass outflow rate and kinetic output with respect to the bolometric and Eddington  luminosity. To this end we need an estimate of  the black hole mass and the   possible launching radius   of the disk wind.

An estimate of the black hole mass was derived   using the   relation between the SMBH mass and the stellar velocity dispersion ($\sigma$) of the host galaxy: log$M_\mathrm{BH}/M_\odot=8.12+4.24\times$log$(\sigma/200\,\rm{km\,s^{-1}})$  (\citealt{Gultekin2009}). 
An  estimate of  the stellar velocity dispersion   can be obtained from the width of the  narrow emission lines, like  [OIII]$\lambda$5007\AA, under the assumption that  the Narrow Line Region gas  is influenced by the potential of the host galaxy  (\citealt{Shields2003}). The  width of the narrow core only of the [OIII]$\lambda$5007\AA\  emission line was measured with  the  public spectrum  of \sorg\ that is  available      in the final release of 6dF
Galaxy Survey (DR3 6dFGS; \citealt{Jones2009}).  Once corrected for the  spectral resolution,  we  measured a  width of $\sigma \sim 185$ km s$^{-1}$, which   corresponds to an estimate of the SMBH mass of $\sim 1\times 10^{8}$\,M$_\odot$.

The bolometric luminosity  ($L_{\mathrm{bol}}$) can be derived  from either the IR luminosity  or from the luminosity of the [OIII]$\lambda$5007\AA\ emission line,  which are   considered    probes of  the accretion disk luminosity.  For the IR luminosity we used the Wide-field Infrared Survey Explorer (WISE) All-Sky catalogue \citep{WISE}. The W3 ($12\,\rm{\mu}$m)  luminosity for \sorg\    is  $L_{\rm WISE,3}=\nu_{\rm W3}\times L_{\rm W3} \sim 1.5\times 10^{44}$\,\lum, where $L_{\rm W3}$\footnote{$L_{\rm W3}$  was obtained from the W3 magnitude $\mbox{w3pro}=5.187$  assuming for a power-law spectrum  ($f_{\nu} \propto \nu^{-\alpha}$) with $\alpha=1$.} and  $\nu_{\rm W3}$ are the monochromatic luminosity and the central frequency  correspondent to the W3 band. 
  We then used the relation between $L_{\rm bol}$ and  $L_{\rm WISE,3}$  derived by \citet{Ballo2014} for a sample  of X-ray selected unabsorbed QSOs, which returned     $L_{\rm bol}\sim1.6\times 10^{45}$\,\lum.   A similar value  ($L_{\rm{bol}} \sim 3\times 10^{45}$\,\lum) is  obtained from   $\lambda L_ {12\,\rm{\mu}\mathrm{m}}$, if we  instead use  the  correlation  derived for a sample of nearby Seyferts (see Fig.5 of \citealt{Gandhi09}).
Finally,  from  the observed  L$_{\rm{[OIII]}}=8.5\times 10^{41}$ \lum reported by \citet{Wu2011} and assuming a bolometric correction of $L_{\rm bol}/L_{\rm {[OIII]}}=3500$ (\citealt{Heckman2004} ), we derived $L_{\rm{bol}} \sim 3.4\times 10^{45}$\,\lum.    Thus,  while uncertain, the above estimates of   $L_{\rm{bol}}=1.6-3.4\times 10^{45}$\,\lum\  and  the SMBH mass ($M_{\mathrm{BH}}\sim 10^{8}\,M_\odot$) suggest that \sorg\ has a moderate accretion rate  ($\lambda_{\mathrm{Edd}}= L_{\mathrm{bol}}/L_{\mathrm{Edd}}=0.12-0.27$), which could be intermediate  between a standard Seyfert and the high accretion rate objects like PDS\,456.

  \subsection{Wind energetics, location and driving mechanisms}
We now explore the main properties of the disk wind detected  in \sorg, focusing at  first on the \suzaku\ observation. For simplicity  we will assume a bolometric luminosity $\sim 3 \times 10^{45}$\,\lum and $M_{\rm BH}\sim 10^{8}\,M_\odot$.   During the 2010 observation two main absorbing zones were found,   both characterised by a high ionisation  (log$\xi= 5.5\pm 0.3$) and a high column density ($\nhsym=5.4-7.8 \times 10^{23}$\,\nh).  
From the  observed outflow velocity of the ionised  gas we  can infer  a lower limit on the launching radius, by equating it to its escape  radius $R_{\rm min}=c^2/v_{\rm out}^2R_{\rm s}$, where $R_{\rm s}\sim 3\times 10^{13}$\,cm is the Schwarzschild radius for \sorg. Thus,  for the two components of the disk wind, we derive $R_{\rm min1}\sim 180\,R_\mathrm{s}$  ($\sim 5.4\times 10^{15}$\,cm) and  $R_{\rm min2}\sim 25\,R_\mathrm{s}$ ($7.5\times10^{14}$\,cm), for the slow  ($v_{\rm {out1}}\sim 0.075\pm -0.01\, c$) and the fast ($v_{\rm {out2}}\sim -0.2\pm 0.02\, c$) component, respectively. 

In order to infer the overall wind  energetics   and possible driving mechanism,  the second main parameter  that we need to quantify  is the mass outflow rate ($\dot M_{\rm out}$).    This can be derived with the equation $\dot M_{\rm {out}}=f\, \pi\,  \mu \, m_{\mathrm p}\,M_{\rm BH}\, v_{\rm {out}}\, R\, N_{\rm {H}}$, which   assumes a biconical geometry for the flow (\citealt{Krongold2007}), where  $\mu=n_{\rm H}/n_{\rm e}=1.4$ for solar abundances and $R$ is the disk wind radius. The parameter $f$ is  a function that  accounts  for the geometry of the system  (i.e. inclination  with respect to the disk and the line of sight). Since  we do not know the exact geometry  for \sorg,  we assumed $f\sim 1.5$, following the same arguments presented in \citet{Tombesi2013}.

For the  first zone (zone1), this yields a mass outflow rate of $\dot M_{\rm{out1}}\sim 1\times 10 ^{26}$\,g s$^{-1}$ ($\sim 1.5 M_\odot$ yr$^{-1}$) and a corresponding kinetic power of  $\dot E_{\rm k1}\sim 2.4\times 10^{44}$\,\lum, which is  approximately  $\sim 8$\% of the bolometric luminosity  (or $\sim 2$\% of  $L_{\rm Edd}$).  Thus, this zone is already   energetically significant  and can provide the feedback mechanisms between the  central SMBH  and the  host galaxy, because it exceeds the theoretical thresholds for feedback  ($\dot E_{\rm {kin1}}/L_{\rm{Bol}}\sim  0.5-5$ \%, \citealt{HopkinsElvis2010,DiMatteo2005}). It is interesting now to compare the outflow momentum rate, $\dot p_{\rm {out1}}= \dot M_{\rm {out1}}\, v_{\rm {out1}}$, with the radiation momentum rate $\dot p_{\rm {rad}}= L_{\rm {bol}}/c$. Although these estimates are  rather uncertain, we found $\dot p_{\rm {out1}}\sim 2\times 10^{35}$\,g cm s$^{-2}$  and  thus of the same order of  $\dot p_{\rm {rad}}$; this suggests  that this zone could be radiation driven with a reasonable  force multiplier.
Although the  fast component of the  disk wind has a similar column density ($\nhsym=5.4 \errUD{3.3}{3.0}\times 10 ^{23}$\, \nh),  the outflow rate is smaller  ($\dot M_{\rm {out2}}\sim 2.6\times 10 ^{25}$\,g s$^{-1}$),  because  the launching  radius  is   smaller compared to zone1.  However, the impact of this possible disk-wind component could be  more important in terms of its feedback.  Indeed, given its   higher velocity ($v_{\rm {out2}}\sim -0.2\, c$), the kinetic power could be of the order of $5\times 10^{44}$\,\lum\ that corresponds to $\sim 15$\% of  $L_{\rm bol}$. 

Regarding  the fastest zone (zone 2)  detected during the \xmm-\nustar\ observation, although the mass outflow rate could be of the same order of the one derived for the fast zone observed with \suzaku, as the launching radius is even smaller ($\sim 8 \,R_\mathrm{s}$), the high velocity implies a high kinetic power ($\sim 40$\% $L_{\rm Bol}$).  However,  all  the  parameters that we can derive  for this zone are highly speculative, because its  $\nhsym $ can be   constrained  only for a given ionisation (see Table~3; $N_{\rm H2}=8\errUD{5}{3}\times 10^{23}$\nh).

  \subsection{The occultation event: evidence for a clumpy disk wind}
We will now discuss the properties of the slow component of the disk wind, which is found in both  of the observations,   within   a clumpy disk wind scenario. First of all, following the same arguments discussed above,   for the slow zone, detected  in the first part of  the \xmm-\nustar\ observation, we derive:  $R_{\rm min}\sim 200\,R_\mathrm{s}$  ($6\times 10^{15}$\,cm),  $\dot E_{\rm k}\sim 6\times 10^{43}$\,\lum\ and  $\dot p_{\rm {out}}\sim 5.5\times 10^{34}$\, g cm s$^{-2}$. 
A possible scenario is that  the disk wind  detected at the beginning of the \xmm-\nustar\ observation     represents the most stable  component of the wind.
The   absorption event that occurred  during sliceB,   can   then be  explained with an increase of  its column density, when a denser region or clump  of the wind moves across the line of sight.  Assuming that its Keplerian (rotation) velocity is similar to the outflow velocity, we can derive the size scale of this denser clump.    From the  observed duration of the occultation  $\Delta t \sim 120 $\,ksec and  assuming $v_{\rm K}=v_{\rm out}=0.13\,c$, we derive  a radial extent of the cloud of  $\Delta R\sim5\times 10^{14}$\, cm  ($\sim 16 \, R_{\rm s}$). Then for   the    observed  column density    variation ($\Delta N_{\rm H}\sim 1.4\times 10^{24}$\,\nh) and  $\Delta R\sim 5\times 10^{14}$\, cm,  the density of the  cloud  is $n_{\rm e}=\Delta  N_{\rm H}/ \Delta R \sim  3\times 10^{9}$\, cm$^{-3}$.  From the    definition  of the ionisation parameter  $\xi=L_{\rm {ion}}/n_{\rm e}R^2$ and assuming that the ionising luminosity  is of the order of $10^{45}$\,\lum (i.e $\sim 1/3 $ of the total bolometric luminosity), we derive a location   for this eclipsing cloud of $R\sim 5\times 10^{15}$\, cm ($\sim 170 R_{\rm s}$) and thus consistent with the location of the bulk of the wind.

\subsection {The Overall scenario of the Wind in \sorg}
\sorg\ is thus a new candidate for an extremely powerful  and stratified disk wind. The two   faster components of the disk wind, seen during the \suzaku\ and the   \xmm\  \& \nustar\ observations  can    be two different inner streamlines of the wind. One possibility is that during the 2015 observation  the inner streamline  of the wind is launched  from closer in.  The slow component of the wind  (zone1) has been detected in both the deep observations, and   it is  outflowing with a velocity of the order of $\sim -0.1\,c$.  

 A possible scenario is that    zone1   is   the more long-lived  component of the wind, which is likely launched from within a few hundreds of $R_{\rm s}$  from the central black hole. Nonetheless, this component is also highly inhomogeneous. Indeed, during the second part of the \nustar\ observation we witnessed an occultation, whereby  our line of sight   intercepted a  clump or filament of the wind,   leading to an increase in column density of $\Delta \nhsym \sim 1.4\times 10^{24}$\,\nh.    At face value the  column  density that is measured in sliceB  would imply a  large $\dot M_{\rm {out}}$  and  a  kinetic output $\dot E_{\rm k}\sim 8 \times 10^{44}$\,\lum,  and  thus more difficult to be  steadily driven.
 As observed this   clump ($\Delta R=16 R_{\rm s}$) corresponds to a short lived ($\Delta t \sim 120 $\,ksec) ejection event. This may  suggests that additional accelerating  mechanisms may be  at work to produce such an ejection, like  magnetic pressure (as  in the MHD wind models; \citealt{Fukumura2010}), which then increase the  momentum of the flow.  We found no evidence that any of these variations could be a response to a different ionising luminosity; indeed, the 2--10 keV luminosities  measured in 2010 and in 2015 are  similar ($L_{2010}\sim 1.4\times 10^{43}$\, \lum\    and $L_{2015}\sim 1.1\times 10^{43}$\, \lum).

 \section{acknowledgements}
 We thank the  referee  Anna Lia Longinotti  for  her  useful comments that  improved  the paper.
This research has made use of data obtained from  \suzaku, a collaborative mission between the space agencies of Japan (JAXA) and the USA (NASA). Based on observations obtained with \xmm, an ESA science mission with instruments and contributions directly funded by ESA Member States and the USA (NASA).
This work made use of data from the \nustar\ mission, a project led by the California Institute of Technology, managed by the Jet Propulsion Laboratory, and funded by NASA. This research has made use of the NuSTAR Data Analysis Software (NuSTARDAS) jointly developed by the ASI Science Data Center and the California Institute of
Technology. VB, GM, PS, AC and RD  acknowledge support from the Italian Space Agency (contract ASI INAF NuSTAR I/037/12/0). VB also  acknowledges  financial  support  through  grants NNX17AC40G and the Chandra grant GO7-18091X.  GM is supported by a European Space Agency (ESA) Research Fellowship. JR  acknowledges  financial  support  through  grants NNX17AC38G, NNX17AD56G and HST-GO-14477.001-A. C.C. acknowledges funding from the European Union's Horizon 2020 research and innovation programme under the Marie Sklodowska-Curie grant agreement No. 664931

 \end{document}